\documentclass{tufte-handout}

\usepackage{enumitem}
\usepackage{amsmath}

\usepackage{graphicx}
\setkeys{Gin}{width=\linewidth,totalheight=\textheight,keepaspectratio}
\graphicspath{{graphics/}}

\usepackage{booktabs}

\usepackage{units}

\usepackage{fancyvrb}
\fvset{fontsize=\normalsize}

\usepackage{multicol}

\usepackage{lipsum}

\usepackage{hyperref} % provides \url{}
\usepackage{enumitem}
\setlist{nosep}

\title{Music Genre Bars}
\author{Swaroop Panda, Shatarupa Thakurta Roy}
\date{Indian Institute of Technology Kanpur}  % if the \date{} command is left out, the current date will be used
\begin{document}

\maketitle% this prints the handout title, author, and date

\begin{abstract}
% \noindent \textsc{Handouts should be made to complement serious presentations.} 
Music Genres, as a popular meta-data of music, are very useful to organize, explore or search music datasets. Soft music genres are weighted multiple-genre annotations to songs. In this initial work, we propose horizontally stacked bar charts to represent a music dataset annotated by these soft music genres. For this purpose, we take an example of a toy dataset consisting of songs labelled with help of three music genres; Blues, Jazz and Country. We demonstrate how such a stacked bar chart can be used as a slider for user-input in an interface. We implement this by embedding this genre bar in a streaming application prototype and show its utility in choosing playlists. We finally conclude by proposing further work and future explorations on our proposed preliminary research.
\end{abstract}

\section{Introduction}
Genres form an important meta-data for music. Music genres like Blues, Jazz, Pop, Rock or Hip-Hop help to identify and locate individual tracks and also to classify, categorize and organize large datasets. A blues song has an distinct and identifiable musical signature as compared to a jazz song. As described in \cite{lena2008classification}, genres organize the production and consumption of cultural material(music in this case). Thus, music genres are used in music classification, recognition, retrieval and recommendation systems. These music genres may be manually annotated by the music artists themselves, by music experts or (with the advent of sophisticated signal processing and machine learning systems) by algorithms.

Visualization of such music datasets is an interesting problem. Visualization helps to better organize and sort, succinctly cluster and improves the overall accessibility of the music dataset. These activities further aids the music consumption and production activities. A lot of work on visualization and organization of music datasets has been done by the music research community. For example, the work in \cite{morchen2005databionic} is a visualization of a music collection according to perceptual distance. Similarly, the work in \cite{lehwark2008visualization} clusters and visualizes tagged music data. \cite{pampalk2001islands} presents a GUI (which can be thought of as a conditioned interactive visualization system) for organization of music archives. The work in \cite{hilliges2006audioradar} presents a digital music player to browse music libraries and generate playlists according to user's mood. All of these works, in one way or the other, are an attempt to organize and sort a music dataset by means of visualization.

Bar charts are ubiquitous on the web, media and in academic research. They are basically used to visually represent quantities associated with a set of related items. They encode quantities by length and are informed by neatly labelled axes and/or legends \cite{streit2014bar}. Over time, different kinds of bar charts have emerged from modifications that have been incorporated into the simple bar chart to accomplish specific visualization goals; examples of which include stacked bar charts, layered bar charts and group bar charts.      

In this preliminary work, we use these bar charts to visualize a music dataset labelled by genres. More specifically, the music dataset contains music files which are annotated by soft music genres. Soft music genres, as we shall shortly come to define, are weighted multiple-genre assignments to songs. The contributions of this initial work include presenting bar charts to represent soft music genres, using bar charts as a slider in an User Interface(UI) and proposing a potential application these genre bars in a streaming application interface. 

\section{Soft Music Genres}
Music genres mostly are annotated according to hard clustering. This means that a song only belongs to the Blues Genre or the Hip-Hop Genre. It cannot belong to both. In contrast, soft clustering allows for the song to belong to two or more clusters (genres in this case) in some proportions. So given a set of three music genres, Jazz, Blues and Country, a song can very well belong to 50\% Jazz, 25\% Blues and 25\% Hip-Hop. The percentages denote a proportion of genre allocation of the particular song. This is basically idea of \textit{soft} music genres. Hard clustering, would have resulted a song being either of Jazz, Blues or Country.
\begin{figure}[htb]
\centerline{
\includegraphics[width=7.5cm]{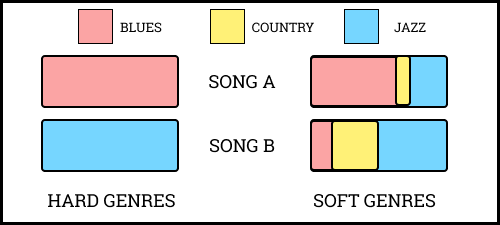}}
  % an empty figure just consisting of the caption lines
  \caption{\label{fig:ex1}A visual depiction of Hard and Soft music genres of two particular songs using bars. The length of the bar is informed by the proportion of the genres present.}
\end{figure}

These soft music genres may be annotated manually or by using algorithms. Soft music genres are most helpful in cases where there expert disagreement over genre allocation; as they allow for doubts or estimates to be incorporated (as in proportions). Allocating genres in proportions is also very helpful while collecting user responses for genre assignment. The advent of sophisticated machine learning and signal processing algorithms has also facilitated the use of soft genre assignment (through soft clustering). An example of such a work includes \cite{panda2019visualizing}. The work uses a topic model on a music genre dataset to obtain probabilistic genres for songs. Topic models are models capable of soft clustering, by using which the authors obtain a soft genre assignment for songs.

\section{Bars as sliders in an UI}
Bar charts, as previously noted, have been and can potentially be used in a lot of different applications. One such application, we propose, could be using stacked bars as sliders in an user interface. Specifically, horizontally stacked bars (with their length commensurate to proportions) can be deployed for enabling the users to interact and modify the proportions. The user can change the length of the bar to adjust the (in this case) music genre proportion. The horizontal stacked bar thus acts just like the popular slider in an UI; allowing users to change values.

\section{Music Genre Bars}
For a preliminary demonstration of music genre bars, say we have a toy music genre dataset, manually labelled by music experts. Music experts conjecture that the files in the dataset belong to only Blues, Country and Jazz genres. The files thus are allocated to 3 music genres in some proportions. To visualize this data we use a simple horizontally stacked bar chart. This bar chart acts as a slider in the UI allowing the user to manipulate values. The music genre bars is stacked of three different sub-bars, with each sub-bar corresponding to a genre and length conditional to the genre proportion. The two intersections of these sub-bars enable the user to change the length of the sub-bar and thus the proportion of the corresponding genres.

% Take dummy data and implement.

% Insert Table here please.

\begin{figure}[htb]
\centerline{
\includegraphics[width=7cm]{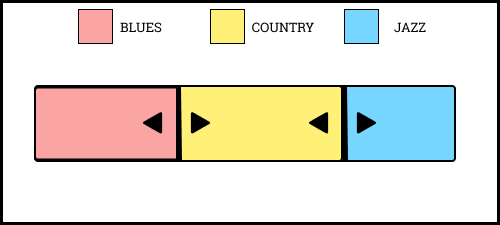}}
  % an empty figure just consisting of the caption lines
  \caption{\label{fig:ex1}The Music Genre Bar}
\end{figure}
% \section{Application}
% Embed in a streaming service platform.
These music genre bars can be used to represent individual songs in a dataset (Figure 1) or the complete dataset. These bars can be used to visually represent an entire genre dataset by means of an underlying search algorithm; as in given a proportion (as displayed on the genre bar), find the songs that represent such proportion. This (as we shall shortly show an instance) can be implemented in a variety of different applications.

\subsection{Searching using the Genre Bars}

The biggest challenge in using such a bar representation is to deal with fact that using bar sliders provides for continuous values that might not be precisely corresponding to genre proportions of the songs. For example, by changing the length of the sub-bars the user may specify the genre proportions to be 22.3\% Blues, 60\% Country and 17.7\% Jazz. And there is no song in a dataset that matches these exact proportions of genres. To get around such this problem, the idea of euclidean distance can be used. Thus, the task changes from of searching for exact proportions within the dataset to searching for a given number of songs that are closest (by euclidean distance) to the proportions provided for by the user. And this idea makes sense in the context of soft music genres, as genres are estimated and proportionally weighted; thus making it more appropriate to present an approximated collection of songs that are close to the given genres than a specific one.       

\section{Implementation}
For an initial implementation, we show how such a horizontally stacked genre bar can be used within a streaming app. We do this using a wireframe prototype and demonstrate how the bars can be combined with playlist (which is just another dataset) selections. Our example (in Figure 3) demonstrates how an input of genre proportions by the user retrieves a selection of songs (a playlist) from a larger database labelled by soft music genres. 
\begin{figure}[htb]
\centerline{
\includegraphics[width=5.5cm, height = 9.5cm]{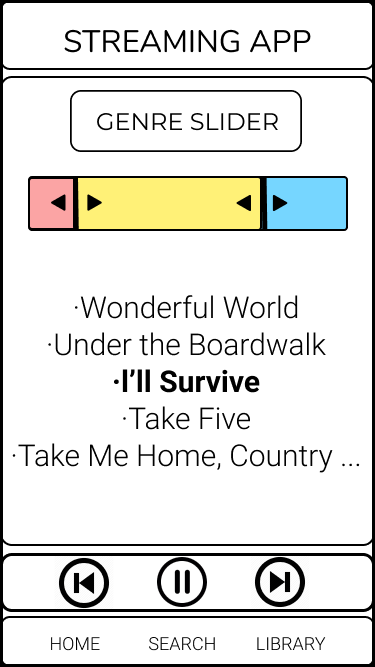}}
  % an empty figure just consisting of the caption lines
  \caption{\label{fig:ex1}The Music Genre Bar embedded as a slider inside a music streaming application. The five songs that appear on the playlist are closest (by euclidean distance) to the proportion as represented by the bar slider.}
\end{figure}
% \section{Evaluation}

\section{Discussion}
In this preliminary work, we presented the concept of music genre bars. Music genre bars represent music files annotated by soft music genres (genres that are proportionately allocated). These genre bars, we suggested can be used as sliders in an UI that an enable user to change values; and thus genre proportions. Genre bars powered by a search algorithm (based on euclidean distance) can represent a dataset. We used a toy example of 3 music genres and demonstrated how these bars can be embedded and deployed in a streaming service platform.  

Further work includes validation and evaluation of these music genre bars. The idea is to evaluate whether users are able to mentally model the sub-bar length with genre proportions. Also, we suppose, it would interesting to understand user perspectives on the idea of soft music genres. The genre bars embedded inside a music streaming UI needs to be rigorously evaluated as well. We suppose, proposed task-based evaluations (on say navigating playlists) could be helpful for this purpose. Visually, the genre-bar needs to adapt to the interface. Future exploration of this work would include incorporating 4 or more genres, larger genre datasets, automatic genre-labelling algorithms and novel methods for evaluation.

\bibliography{sample-handout}

\begin{thebibliography}{7}
\providecommand{\natexlab}[1]{#1}
\providecommand{\url}[1]{\texttt{#1}}
\expandafter\ifx\csname urlstyle\endcsname\relax
  \providecommand{\doi}[1]{doi: #1}\else
  \providecommand{\doi}{doi: \begingroup \urlstyle{rm}\Url}\fi

\bibitem[Hilliges et~al.(2006)Hilliges, Holzer, Kl{\"u}ber, and
  Butz]{hilliges2006audioradar}
Otmar Hilliges, Phillipp Holzer, Rene Kl{\"u}ber, and Andreas Butz.
\newblock Audioradar: A metaphorical visualization for the navigation of large
  music collections.
\newblock In \emph{International Symposium on Smart Graphics}, pages 82--92.
  Springer, 2006.

\bibitem[Lehwark et~al.(2008)Lehwark, Risi, and
  Ultsch]{lehwark2008visualization}
Pascal Lehwark, Sebastian Risi, and Alfred Ultsch.
\newblock Visualization and clustering of tagged music data.
\newblock In \emph{Data Analysis, Machine Learning and Applications}, pages
  673--680. Springer, 2008.

\bibitem[Lena and Peterson(2008)]{lena2008classification}
Jennifer~C Lena and Richard~A Peterson.
\newblock Classification as culture: Types and trajectories of music genres.
\newblock \emph{American sociological review}, 73\penalty0 (5):\penalty0
  697--718, 2008.

\bibitem[M{\"o}rchen et~al.(2005)M{\"o}rchen, Ultsch, N{\"o}cker, and
  Stamm]{morchen2005databionic}
Fabian M{\"o}rchen, Alfred Ultsch, Mario N{\"o}cker, and Christian Stamm.
\newblock Databionic visualization of music collections according to perceptual
  distance.
\newblock In \emph{ISMIR}, pages 396--403, 2005.

\bibitem[Pampalk(2001)]{pampalk2001islands}
Elias Pampalk.
\newblock \emph{Islands of music: Analysis, organization, and visualization of
  music archives}.
\newblock na, 2001.

\bibitem[Panda et~al.(2019)Panda, Namboodiri, and Roy]{panda2019visualizing}
Swaroop Panda, Vinay~P Namboodiri, and Shatarupa~Thakurta Roy.
\newblock Visualizing music genres using a topic model.
\newblock 2019.

\bibitem[Streit and Gehlenborg(2014)]{streit2014bar}
Marc Streit and Nils Gehlenborg.
\newblock Bar charts and box plots, 2014.

\end{thebibliography}
\bibliographystyle{plainnat}

\end{document}